\documentclass[12pt]{article}
\usepackage{amsmath}
\usepackage{amssymb}
\usepackage{graphicx}

\newcommand{\rmd}{{\mathrm d}}
\newcommand{\Or}{{\mathrm O}}

\begin{document}

\title{Unstable periodic orbits in a chaotic meandering jet flow}
\author{M. Yu. Uleysky, M. V. Budyansky, S. V. Prants}
\maketitle

\begin{abstract}
We study the origin and bifurcations of typical classes of unstable
periodic orbits in a jet flow that was introduced before as a kinematic
model of chaotic advection, transport and mixing of passive scalars
in meandering oceanic and atmospheric currents. A method to detect
and locate the unstable periodic orbits and classify them by the origin
and bifurcations is developed. We consider in detail period-1 and period-4
orbits playing an important role in chaotic advection. We introduce five
classes of period-4 orbits: western and eastern ballistic ones, whose
origin is associated with ballistic resonances of the fourth order,
rotational ones, associated with rotational resonances of the second
and fourth orders, and rotational-ballistic ones associated with
a rotational-ballistic resonance. It is a new kind of nonlinear resonances
that may occur in chaotic flow with jets and/or circulation cells.
Varying the perturbation amplitude, we
track out the origin and bifurcations of the orbits
for each class.
\end{abstract}

\section{Introduction}
It is well known that a dynamical system is chaotic if it
displays sensitivity to initial conditions, has a dense orbit
and a dense set of periodic orbits (see, for example, \cite{Holmes}).
Periodic orbits play an important role in organizing
dynamical chaos both in Hamiltonian and dissipative systems.
Stable periodic orbits (SPO) organize a regular motion inside
islands of stability in the phase space. Unstable periodic orbits (UPO)
form a skeleton around which chaotic dynamics is organized.
The motion nearby an UPO is governed by its
stable and unstable manifolds. Owing to the density property,
the UPOs influence even the asymptotic dynamics. Order
and disorder in a chaotic regime are produced eventually
by an interplay between sensitivity to initial conditions
and regularity of the periodic motion.

In the present paper we study origin and bifurcations of typical
classes of the UPOs in a two-dimensional incompressible flow
that has been introduced and analyzed in Refs.~\cite{S92,DW96,PBUZ06,
SamWig2006,UBP07} as a toy
kinematic model of transport and mixing of passive particles
in meandering jet currents in the ocean, like the Gulf Stream, Kuroshio
and other main oceanic currents, and in the atmospheric currents
like zonal jets with propagating Rossby waves
\cite{WK89, P91, DM93}. The equations of motion
of passive particles advected by any incompressible planar
flow are known to have a Hamiltonian form
\begin{equation}
\frac{\rmd x}{\rmd t}=u(x,y,t)=-\frac{\partial\Psi}{\partial y},\quad
\frac{\rmd y}{\rmd t}=v(x,y,t)=\frac{\partial\Psi}{\partial x},
\label{adveq}
\end{equation}
where the streamfunction $\Psi$ plays the role of a Hamiltonian, and
the particle's coordinates $x$ and $y$ are canonically
conjugated variables. The phase space of Eqs.~\ref{adveq}
is a physical space for advected particles. If the
velocity field, $u=u(x,y)$ and $v=v(x,y)$, is stationary,
then fluid particles move along streamlines, and the motion is
completely regular with any Eulerian stationary field whatever
its complexity. A time-periodic velocity field,
$u(x,y,t)=u(x,y,t+T)$ and $v(x,y,t)=v(x,y,t+T)$, can produce chaotic
particle's trajectories, the phenomenon known as
``chaotic advection''~\cite{A02, O89}.

\begin{figure}[!tpb]
\centerline{\includegraphics[width=0.7\textwidth]{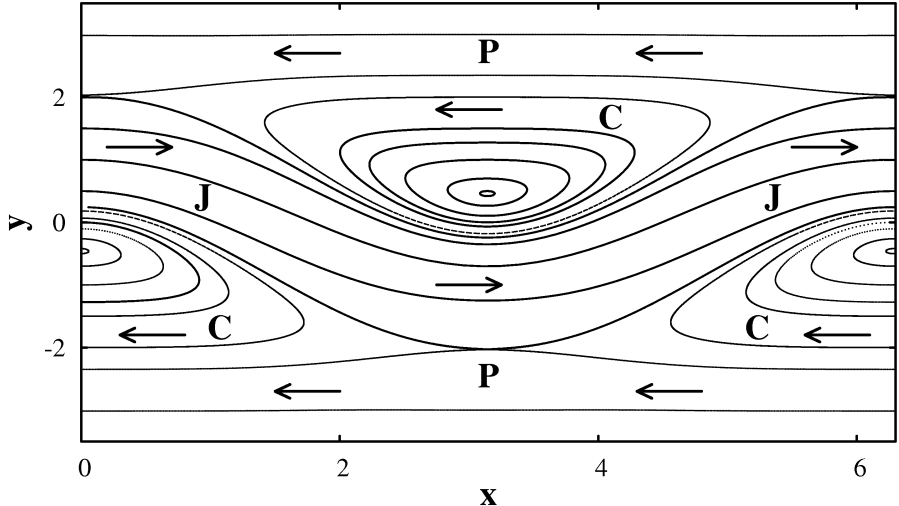}}
\caption{The phase portrait of the model flow (\ref{psim}) in the
comoving frame of reference. The first frame with streamlines
in the circulation (C), jet (J) and peripheral currents (P)
zones is shown. The parameters are: $A=0.785$, $C=0.1168$ and $L=0.628$.}
\label{phasespace}
\end{figure}

Chaotic advection of water (air) masses along with their properties
in geophysical jets is a topic of great interest in the last
decade (for recent reviews on chaotic advection, transport and mixing
in the ocean and atmosphere see \cite{KP06} and \cite{HPS07},
respectively).
Among the variety of kinematic and dynamic models of shear flows,
one of the simplest ones is a Bickley jet with the velocity
profile $\sim\mathop{\mathrm{sech}}^2 y$ and a running wave imposed.
The phase portrait of such a flow in the frame, moving with the
phase velocity of the running wave, is shown in Fig.~\ref{phasespace}.
The flow consists of three distinct regions, the eastward jet (J),
the circulations (C) and the westward peripheral currents (P) to the
north and south from the jet, separated from each other by
the northern and southern $\infty$-like separatrices.
A simple periodic modulation of the wave's amplitude breaks up
these separatrices, produces stochastic layers in place of them, and
chaotic mixing and transport of passive particles may occur.

In the recent papers \cite{PBUZ06, UBP07} we have studied statistical
properties of chaotic mixing and transport in such a time-periodic
meandering Bickley-jet current and explained some of them by the
presence of dynamical traps in the phase space, singular zones
in the stochastic layers where particles may spend arbitrary long but
finite time \cite{Z02}. We identified rotational-islands traps around
the boundaries of rotational islands, ballistic-islands traps,
around the boundaries of ballistic islands, and saddle traps
associated with stable manifolds of periodic saddle trajectories \cite{UBP07}.

A further insight into chaotic advection in the chosen model flow and in
other jet flows is required in order to find a connection
between dynamical and topological properties of particle's trajectories
forming a complex picture of chaotic mixing. The aim of this paper
is to study in detail origin, properties and bifurcations of typical
UPOs in the flow considered in Refs.~\cite{PBUZ06,UBP07}.
In Sec.~2 we introduce the model streamfunction and advection equations and
present a numerical method for locating the UPOs of different periods in
chaotic dynamical systems. The method is based on computing a distance $d$
between the positions, $x(t_0)$, $y(t_0)$, of a chosen particle at the moments of time
$t_0$ and $t_0+mT_0$ (where $T_0$ is a period of the perturbation and
$m=1,2,\dots$), finding local minima of the distance function with given values of
$m$ and analyzing them to locate the period-$m$ UPOs for which
$d(x(t_0)$, $y(t_0))=0$. In Sec.~3 we analyze a saddle orbit (SO) by linearizing
the advection equations and study its metamorphoses varying the
perturbation amplitude $\varepsilon$. In Sec.~4 we apply the method
to study the origin, properties and bifurcations of the period-4 UPOs.
We chose namely that period because for sufficiently large values
of $\varepsilon$ there are no visible period-4 resonances in the
phase space. It is easy to locate the UPOs and SPOs of visible
resonances on Poincar{\'e} sections, but it is not a trivial job
to do that with broken resonances. All the period-4 UPOs are
classified in 5 distinct groups which differ by the type of motion of the
corresponding particles and their origin and bifurcations.
Varying the  perturbation amplitude $\varepsilon$, we search for those
resonances in each class which generate the corresponding period-4 UPOs
and compute bifurcation diagrams for them.

\section{Return maps for the model flow}
Following to Refs.~\cite{PBUZ06, UBP07}, we consider the following
streamfunction in the laboratory frame of reference:
\begin{equation}
\Psi'(x',y',t')=-\Psi'_0\tanh{\left(\frac{y'-a\cos{k(x'-c t')}}
{\lambda\sqrt{1+k^2 a^2\sin^2{k(x'-c t')}}} \right)},
\label{psil}
\end{equation}
where the hyperbolic tangent produces the Bickley-jet profile,
the square root provides a constant width of the jet
$\lambda$, and $a$, $k$ and $c$ are amplitude, wave number
and phase velocity of the running wave, respectively.
The normalized streamfunction in the frame moving
with $c$ is
\begin{equation}
\Psi=-\tanh{\left(\frac{y-A\cos x}{L\sqrt{1+A^2\sin^2 x}}\right)}+Cy,
\label{psim}
\end{equation}
where $x=k(x'-ct')$ and $y=ky'$ are new scaled coordinates.
The normalized jet's width $L=\lambda k$, wave's amplitude
$A=ak$ and phase velocity $C=c/\Psi'_0 k$ are the control parameters.
The advection equations (\ref{adveq}) with the streamfunction (\ref{psim})
have the following form in the comoving frame:
\begin{equation}
\begin{gathered}
\dot x=\frac{1}{L\sqrt{1+A^2\sin^2 x}{\cosh^2\theta}}-C,\quad
\dot y=-\frac{A\sin x(1+A^2-Ay\cos x)}{L\left(1+A^2\sin^2 x\right)^{3/2}
{\cosh^2\theta}},r\\
\theta=\frac{y-A\cos x}{L\sqrt{1+A^2\sin^2 x}},
\end{gathered}
\label{eqs}
\end{equation}
where dot denotes differentiation with respect to the scaled time
$t=\Psi'_0 k^2 t'$.

The flow with the streamfunction (\ref{psil}) is steady in
the comoving frame and its phase portrait is shown in
Fig.~\ref{phasespace}. There are southern and northern
sets of elliptic fixed points:
$x_e^{(s)}=2\pi n$, $y_e^{(s)}=-L\mathop{\mathrm{Arcosh}}\sqrt{1/LC}+A$ and
$x_e^{(n)}=(2n+1)\pi$, $y_e^{(n)}=L\mathop{\mathrm{Arcosh}}\sqrt{1/LC}-A$,
respectively, and the southern and northern sets of
hyperbolic (saddle) fixed points:
$x_s^{(s)}=(2n+1)\pi$, $y_s^{(s)}=-L\mathop{\mathrm{Arcosh}}\sqrt{1/LC}-A$ and
$x_s^{(n)}=2\pi n$, $y_s^{(n)}=L\mathop{\mathrm{Arcosh}}\sqrt{1/LC}+A$,
respectively, where $n=0,\pm 1, \dots$.

A perturbation is provided by a periodic modulation of the wave's amplitude
\begin{equation}
A(t)=A_0+\varepsilon\cos(\omega t+\varphi).
\label{perturb}
\end{equation}
The equations of motion (\ref{eqs}) are symmetric under the following
transformations: $t\to t$, $x\to\pi+x$, $y\to -y$ and $t\to -t$, $x\to\ -x$, $y\to y$.
Due to these symmetries, the motion can be considered
in the northern chain of the circulation cells on the cylinder with
$0\le x\le 2\pi$. The part of the phase space with
$2\pi n\le x\le 2\pi (n+1)$, $n=0, \pm 1, \dots$, is called {\it a frame}.
The first frame is shown in Fig.~\ref{phasespace}.
The values of the following control parameters are fixed
in  our simulation: $L=0.628$, $A_0=0.785$, $C=0.1168$, $T_0=2\pi/\omega=24.7752$,
$\varphi=\pi/2$. The only varying parameter is the perturbation amplitude
$\varepsilon$.

In fluid mechanics an infinite number of initial conditions comes in play
simultaneously and a number of fluid elements, launched in different places,
may follow the same orbit on the flow plane.
Essentially, that a number of particles with different initial positions may
move along the same orbit. More precisely, {\it an orbit}
is a set of points ${x_i, y_i}$ $(i=1,2,\dots)$ on the phase plane
(on the flow plane) with the following two properties: (i) there exists
for $\forall i, j$ an integer $k$ (positive or negative) such that
\begin{equation}
\left(
\begin{array}{c}
x_i \\ y_i
\end{array}
\right)=\hat U(kT_0)
\left(
\begin{array}{c}
x_j \\ y_j
\end{array}
\right),
\label{orbit}
\end{equation}
where $\hat U$ is an evolution operator; and (ii) there exists
for $\forall i, k$ an integer $j$ such that
\begin{equation}
\hat U(kT_0)
\left(
\begin{array}{c}
x_i \\ y_i
\end{array}
\right)=\left(
\begin{array}{c}
x_j \\ y_j
\end{array}
\right).
\label{orbit2}
\end{equation}
{\it A period-$m$ orbit} is a finite set of points on the phase plane with the
properties (\ref{orbit}) and (\ref{orbit2}) consisting of $m$ elements.
Thus, any period-$m$ orbit contains $m$ points whose trajectories
belong to this orbit.

To locate the UPOs in the phase space we fix values of the control parameters
and compute with a large number of particles
the Euclidean distance
$d^2=[x(t_0+mT_0)-x(t_0)]^2+[y(t_0+mT_0)-y(t_0)]^2$ between
particle's position at an initial moment of time $t_0$ and
at the moments of time $T=mT_0$, where $m=1,2,\dots$.
The data are plotted as {\it a period-$m$ return map} (RM)
that shows by color the values of $d$ for particles with initial
positions $[x(t_0), y(t_0)]$. At the first stage, we select a large
number of points where the function $d(x(t_0),y(t_0))$ may have
local minima. Then we apply the method of a deformed simplex to
localize the minima in neighbourhoods of those points. There are
such minima among them for which $d=0$ with a given value of $m$.
The procedure allows to detect
both UPOs and SPOs not only in periodically perturbed Hamiltonian
systems but in any chaotic system.

\begin{figure}[!tpb]
\centerline{\includegraphics[width=0.49\textwidth]{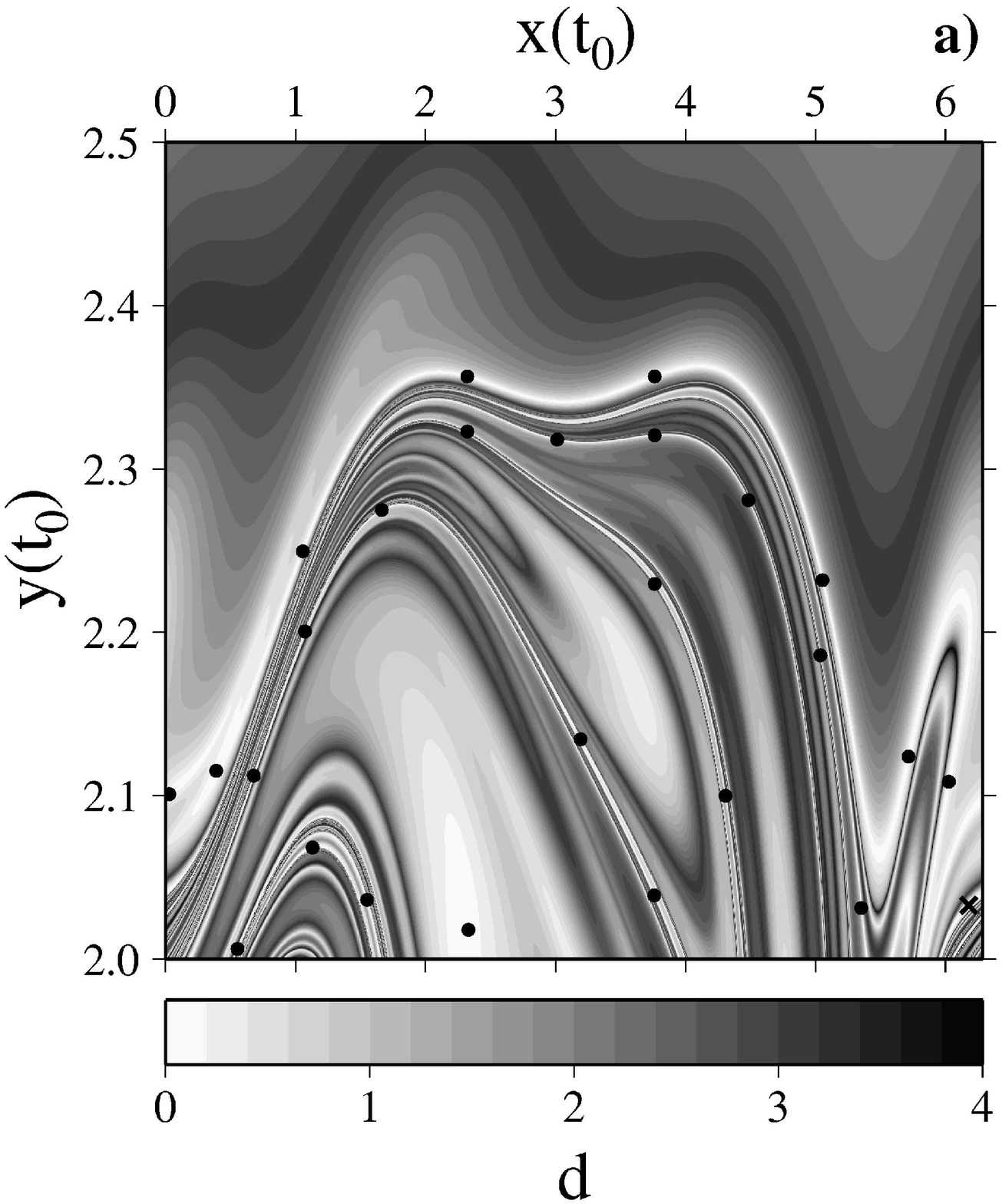}
\includegraphics[width=0.49\textwidth]{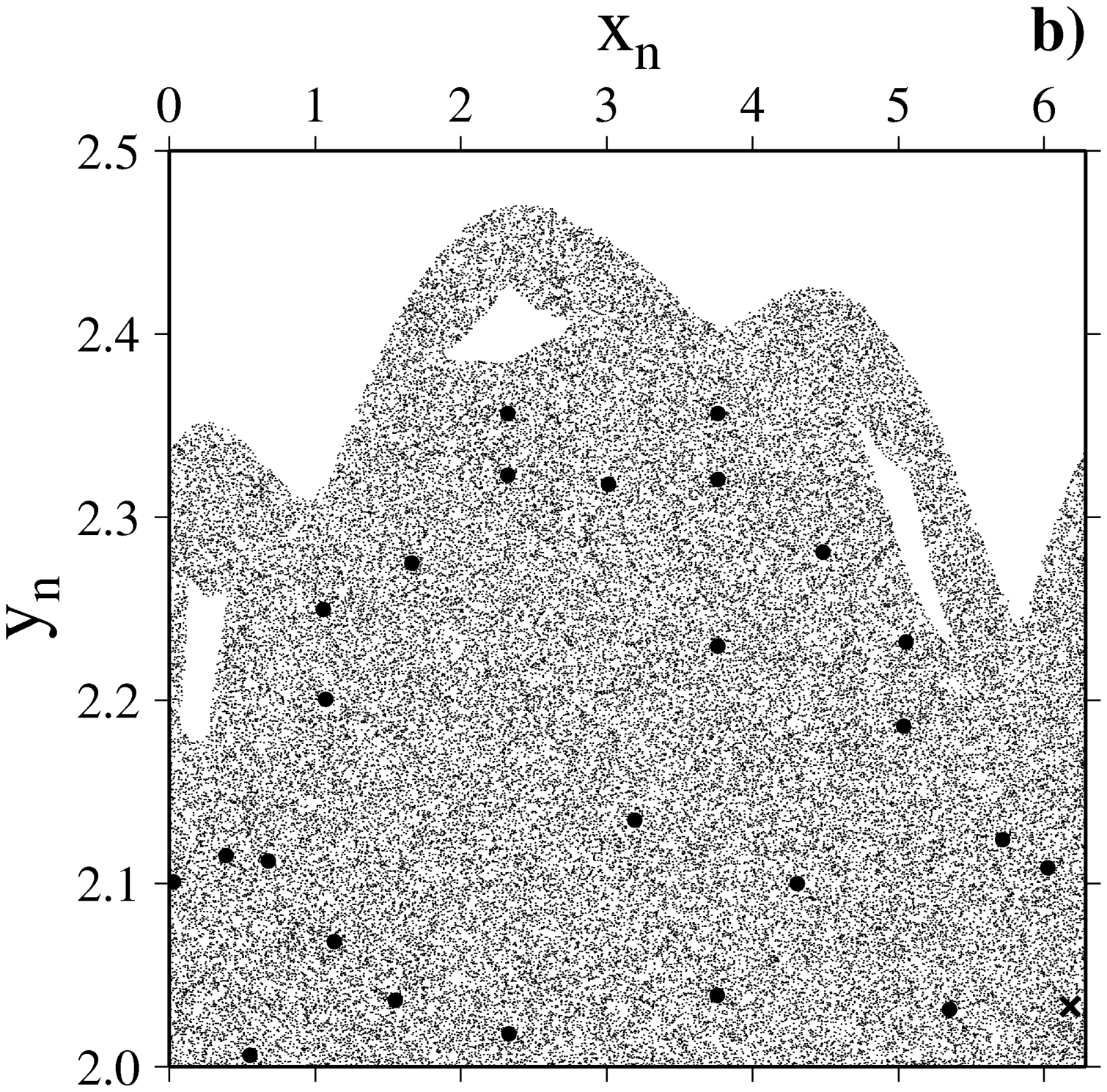}}
\caption{(a) Period-4 return map representing the distance $d$ between
particle's position $x(t_0)$ and $y(t_0)$ at $t_0$ and its position
at $t_0+4T_0$. Cross marks location of the saddle trajectory and dots mark
initial positions of trajectories of the period-4 UPOs.
(b) Poincar{\'e} section of the
northern separatrix layer with positions of those dots.}
\label{Rmap}
\end{figure}

Return maps with $m=1,2,3,4$ and $12$ have been computed. In the next
section we consider the SO which is a period-1 UPO. The main efforts are
devoted to analysis of period-4 UPOs because it is not a trivial task to
detect the UPOs when the corresponding resonances cannot be identified
on Poincar{\'e} sections. It is the case with $m=4$. The period-4 RM is
shown in Fig.~\ref{Rmap}a with the cross marking location of
the period-1 saddle trajectory and black dots marking initial positions
of twenty six trajectories of the period-4 UPOs with a relative
accuracy $10^{-13}\div 10^{-14}$.
For comparison, we demonstrate in Fig.~\ref{Rmap}b
locations of those dots on the Poincar{\'e} section. It is evident that
they cannot be prescribed to any structures in the phase space and
cannot be identified by inspection of the Poincar{\'e} section.

\section{{\bf Saddle orbit}}
The saddle points of the unperturbed equations of motion
(\ref{eqs}), $(x_s^{(n)}, y_s^{(n)})$ and $(x_s^{(s)}, y_s^{(s)})$,
become period-1 saddle orbits (SO) under the periodic perturbation
(\ref{perturb}). To analyze such a SO we linearize the perturbed equations
\begin{equation}
\dot x=X(x,y),\quad \dot y=Y(x,y)
\label{maineqs}
\end{equation}
in a neighborhood of the saddle point
\begin{equation}
x_s=0,\quad y_s=L\mathop{\mathrm{Arcosh}}\sqrt{\frac{1}{LC}}+A_0,
\label{saddlepoint}
\end{equation}
where $X(x,y)$ and $Y(x,y)$ are the right-hand sides
of the corresponding equations in the set (\ref{eqs}) with
$A$ being the time-periodic amplitude (\ref{perturb}).
After linearizing, we get the following equations for small
deviations $\eta\equiv x-x_s$ and $\xi\equiv y-y_s$:
\begin{equation}
\dot \eta=\frac{1}{L\cosh^2\Theta}-C-\frac{2\tanh\Theta}{L^2\cosh^2\Theta}\xi,\quad
\dot \xi=\frac{A(AL\Theta-1)}{L\cosh^2\Theta}\eta,
\label{dots_lin_gen}
\end{equation}
where
\begin{displaymath}
\Theta\equiv\mathop{\mathrm{Arcosh}}\sqrt{\frac{1}{LC}}-\frac{\varepsilon \cos \Phi}{L},\quad
\Phi\equiv \omega t+\varphi.
\end{displaymath}

For small values of the perturbation amplitude $\varepsilon$,
it is possible to simplify the set (\ref{dots_lin_gen}) expanding the
right-hand sides in a series in powers of $\varepsilon\cos\Phi$
and neglecting terms above the first order. In terms of the
variables
\begin{equation}
\begin{gathered}
X_s=\lim_{\varepsilon\to 0}X(x_s,y_s),\quad
X_{ys}=\lim_{\varepsilon\to 0}X_y(x_s,y_s),\quad
Y_{xs}=\lim_{\varepsilon\to 0}Y_x(x_s,y_s),\\
X_\varepsilon=\lim_{\varepsilon\to 0}\frac{\partial X(x_s,y_s)}{\partial (\varepsilon\cos\Phi)},\quad
X_{y\varepsilon}=\lim_{\varepsilon\to 0}\frac{\partial X_y(x_s,y_s)}{\partial (\varepsilon\cos\Phi)},\quad
Y_{x\varepsilon}=\lim_{\varepsilon\to 0}\frac{\partial Y_x(x_s,y_s)}{\partial (\varepsilon\cos\Phi)},\\
X_x(x,y)=\frac{\partial X(x,y)}{\partial x},\quad
X_y(x,y)=\frac{\partial X(x,y)}{\partial y},\\
Y_x(x,y)=\frac{\partial Y(x,y)}{\partial x},\quad
Y_y(x,y)=\frac{\partial Y(x,y)}{\partial y}
\end{gathered}
\label{defs_xyeps}
\end{equation}
we get the following equations of the first order in $\varepsilon\cos\Phi$:
\begin{equation}
\dot \eta=X_s+X_{ys}\xi+\varepsilon\cos\Phi(X_\varepsilon+X_{y\varepsilon}\xi),\quad
\dot \xi=(Y_{xs}+\varepsilon\cos\Phi Y_{x\varepsilon})\eta,
\label{lineqs1}
\end{equation}
where
\begin{equation}
\begin{gathered}
X_s=0,\quad
X_{ys}=-\frac{2C\sqrt{1-LC}}{L},\quad
Y_{xs}=A_0C\left(A_0L\mathop{\mathrm{Arsech}}\sqrt{LC}-1\right),\\
X_\varepsilon=\frac{2C\sqrt{1-LC}}{L},\quad
X_{y\varepsilon}=\frac{2C(3LC-2)}{L^2},\\
Y_{x\varepsilon}=\frac{C\left(2A_0L\left(L+A_0\sqrt{1-LC}\right)\mathop{\mathrm{Arsech}}\sqrt{LC}-2A_0\sqrt{1-LC}-L(1+A_0^2)\right)}{L}.
\end{gathered}
\end{equation}

Coming back to set (\ref{dots_lin_gen}) and expanding the equations up to the
second order in $\varepsilon\cos\Phi$, we obtain the equations
\begin{equation}
\begin{aligned}
\dot\eta&=X_s+X_{ys}\xi+\varepsilon\cos\Phi(X_\varepsilon+X_{y\varepsilon}\xi)+\frac{\varepsilon^2}{2}\cos^2\Phi(X_{\varepsilon^2}+X_{y\varepsilon^2}\xi),\\
\dot \xi&=(Y_{xs}+\varepsilon\cos\Phi Y_{x\varepsilon}+\frac{\varepsilon^2}{2}\cos^2\Phi Y_{x\varepsilon^2})\eta,
\end{aligned}
\label{lineqs2}
\end{equation}
with the following notations:
\begin{equation}
\begin{gathered}
\begin{aligned}
Y_{x\varepsilon^2}=&\lim_{\varepsilon\to 0}\frac{\partial^2 Y_x(x_s,y_s)}{\partial (\varepsilon\cos\Phi)^2}=
-\frac{2C}{L^2}\left[A_0(2L^2-3LC+2)+2L(1+A_0^2)\sqrt{1-LC}-\right.\\
&\qquad\left.-L\left(
L^2+A_0^2(2-3LC)+4A_0L\sqrt{1-LC}
\right)\mathop{\mathrm{Arsech}}\sqrt{LC}\right],
\end{aligned}\\
X_{\varepsilon^2}=\lim_{\varepsilon\to 0}\frac{\partial^2 X(x_s,y_s)}{\partial (\varepsilon\cos\Phi)^2}=
\frac{2C(2-3LC)}{L^2},\\
X_{y\varepsilon^2}=\lim_{\varepsilon\to 0}\frac{\partial^2 X_y(x_s,y_s)}{\partial (\varepsilon\cos\Phi)^2}=
\frac{8C\sqrt{1-LC}(3LC-1)}{L^3}.
\end{gathered}
\label{Coeff_lin_quad}
\end{equation}

\begin{figure}[!tpb]
\centerline{\includegraphics[width=0.49\textwidth]{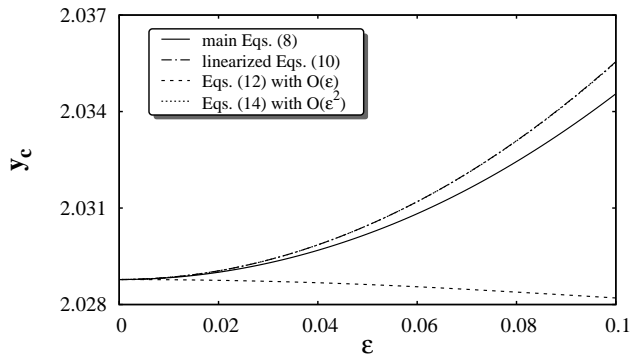}}
\caption{Dependence of the $y$-coordinate of the central
point $y_c$ of the SO on the perturbation amplitude $\varepsilon$
with different versions of advection equations.}
\label{SOcenter}
\end{figure}

\begin{figure}[!tpb]
\centerline{\includegraphics[width=0.49\textwidth]{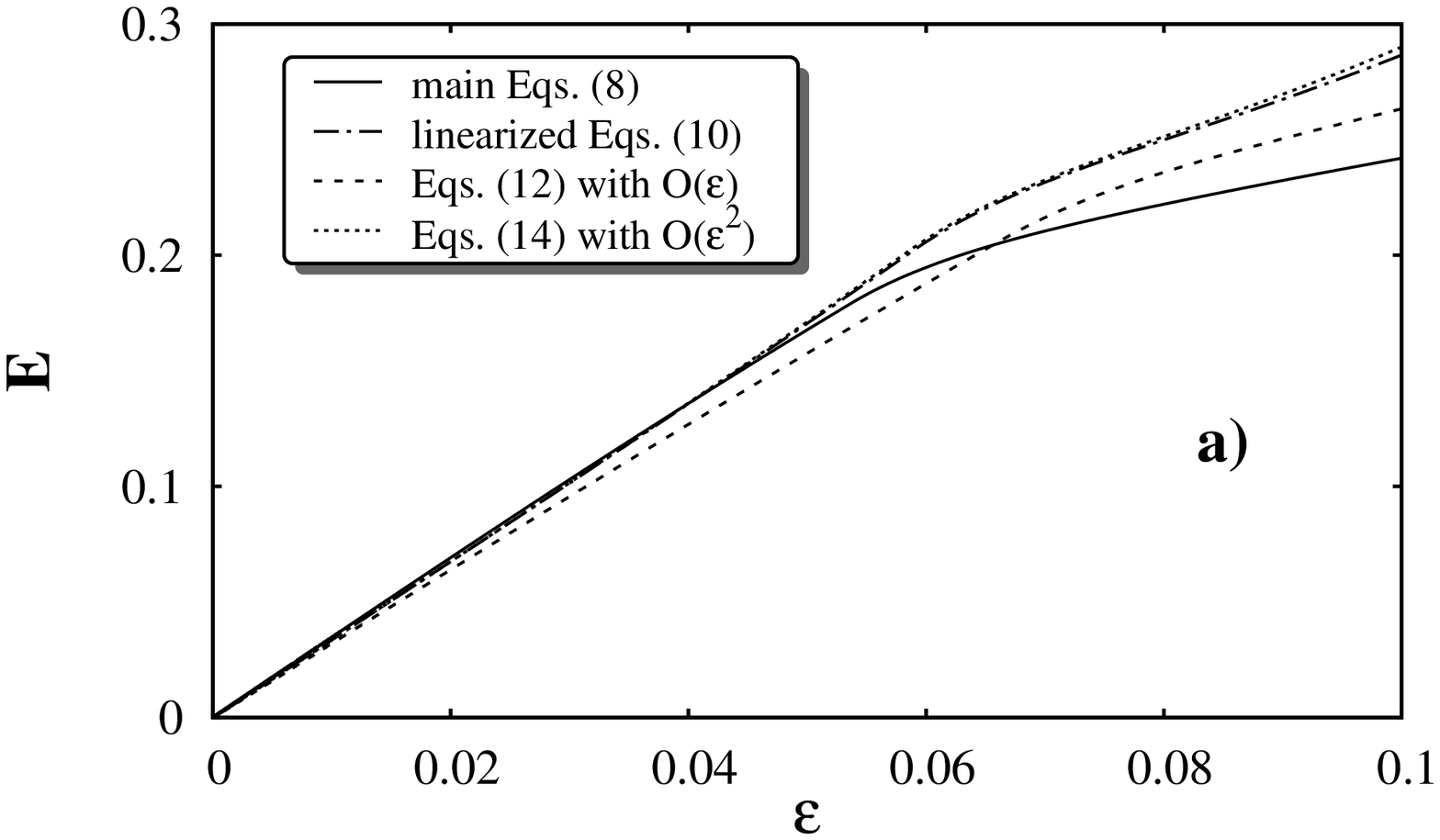}
\includegraphics[width=0.49\textwidth]{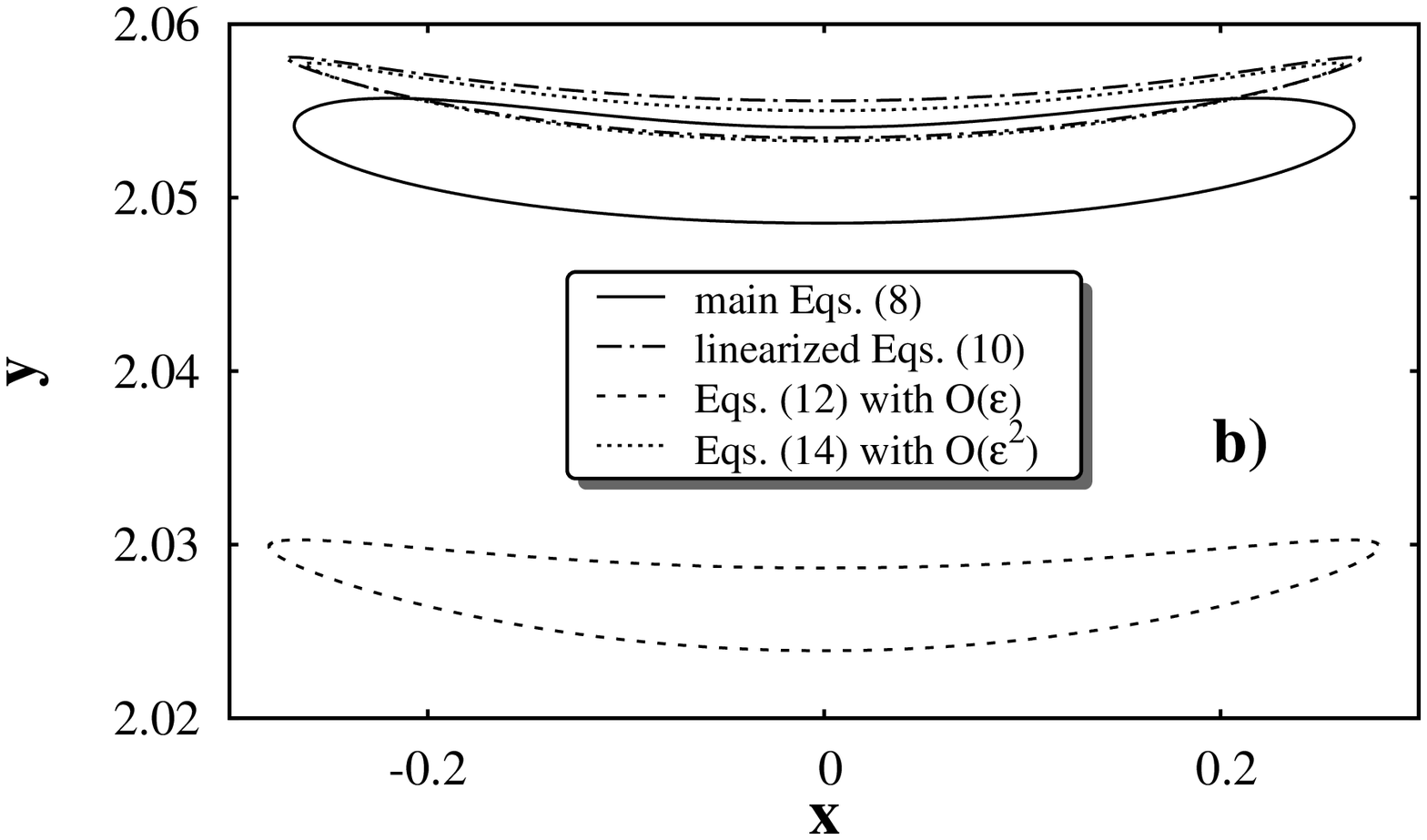}}
\caption{(a) Dependence of deviation of the SO form from ellipticity $E$
on the perturbation amplitude $\varepsilon$ and (b) the form
of the SO at $\varepsilon=0.2$.}
\label{SOform}
\end{figure}

In the following, we compare numerically some properties of the SOs,
generated from the same fixed point $(x_s,y_s)$, with the main
equations of motion (\ref{maineqs}), with linearised equations
(\ref{dots_lin_gen}), with linearized Eqs.~\ref{lineqs1} with $\Or(\varepsilon)$
and Eqs.~\ref{lineqs2} with $\Or(\varepsilon^2)$. Simulation shows that dependence of
$x$-coordinate of the initial position of the SO on the perturbation amplitude
$\varepsilon$ is practically the same with all the versions of the
advection equations listed above.
As $\varepsilon$ increases, the $x$-coordinate shifts to the west
from the point $x_s=0$.
As to $y$-coordinate, it moves to the north with increasing
$\varepsilon$ in a similar way for Eqs.~\ref{maineqs}, \ref{dots_lin_gen} and
\ref{lineqs2}, but its behavior differs strongly for the
first-order equations in $\varepsilon$ (\ref{lineqs1}).
In Fig.~\ref{SOcenter} we demonstrate the $\varepsilon$-dependence
of the $y$-coordinate of the central point $y_c$ of the SO which is a
central point between maximal and minimal values of $y$.
In all the cases, $x_c=0$ is $x$-coordinate of the central point.
It is evident that equations of the first order in $\varepsilon$
cannot give a correct position of the SO. The $\varepsilon$-dependence
of deviation of the form of the SO from ellipticity $E$ is shown in Fig.~\ref{SOform}a,
where $E$ is the ratio of the area between the SO and an ellipse with semi-axes
equal to the height and width of the SO to the area of the ellipse.
Fig.~\ref{SOform}b demonstrates that the form of the SO does not
change dramatically after linearization, but its position
in the phase space cannot be found correctly in the first order in
$\varepsilon$. We may conclude that even with small values of the perturbation
amplitude the effect of the second harmonic $2\omega$ is not small.

\section{Origin and bifurcations of unstable periodic orbits}
In this section we analyze the origin of the period-4 UPOs
and their bifurcations that occur with changing the perturbation
amplitude $\varepsilon$. Using the period-4 RM (Fig.~\ref{Rmap}a),
we located all the period-4 UPOs and the initial positions
of four trajectories for each of them. They differ by the type
of motion of the passive particles and their length $l$ which
is a length of the corresponding curve between two positions on the
orbit separated by the time interval $4T_0$. On the cylinder
$0\le x\le 2\pi$, all the UPOs are closed curves whose topology
may be very complicated. In the physical space, the UPOs can be
classified as rotational ones (particles rotate in the same
frame), ballistic ones (particles move ballistically
from a frame to frame) and rotation-ballistic ones
(particles may rotate for a while in a frame, then move ballistically
through a few frames and change their direction of motion).

\subsection{$C_\mathrm{WB}^{4:1}$ class: western ballistic UPOs
associated with the $4:1$ western ballistic resonance}

\begin{figure}[!tpb]
\centerline{\includegraphics[width=0.49\textwidth]{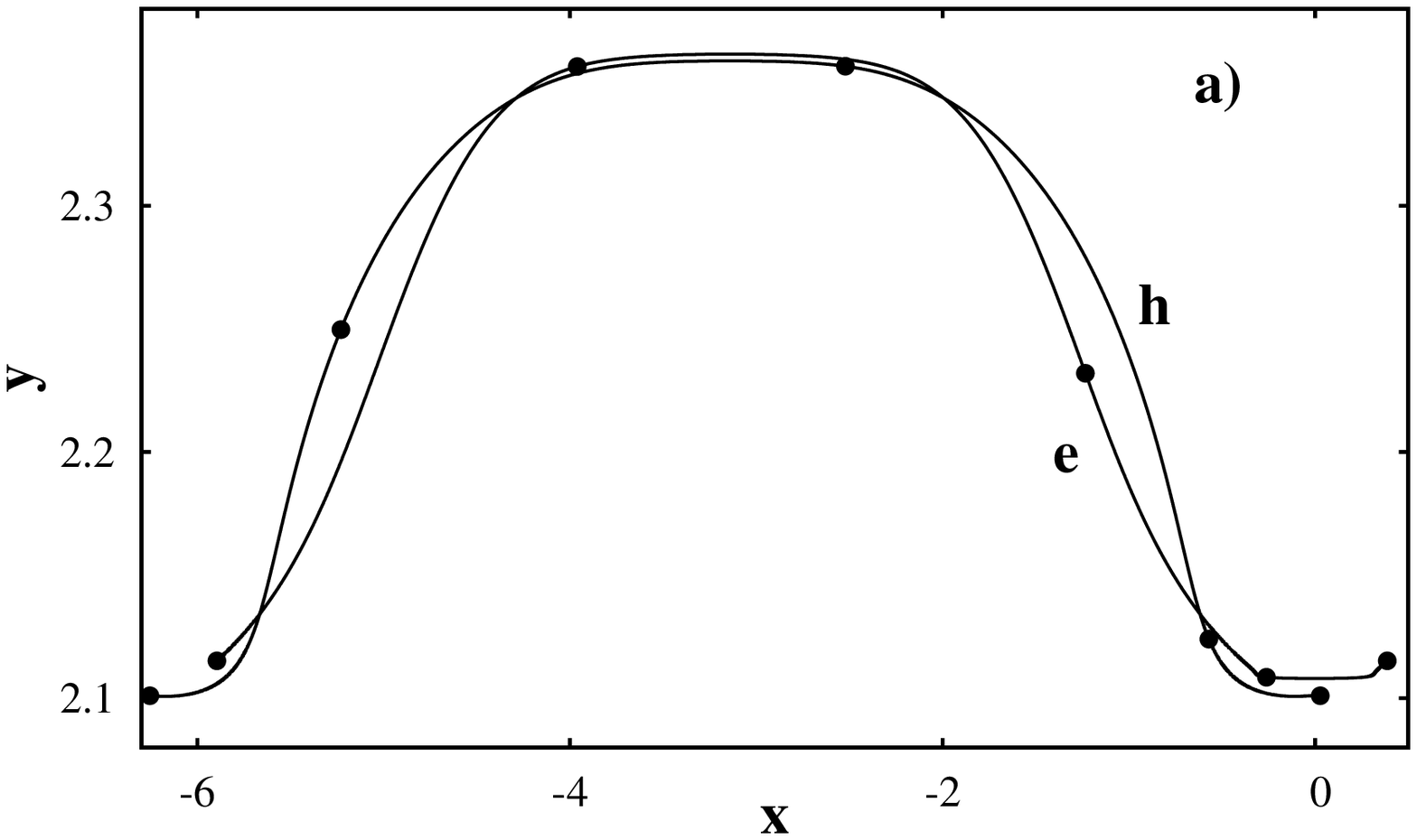}
\includegraphics[width=0.49\textwidth]{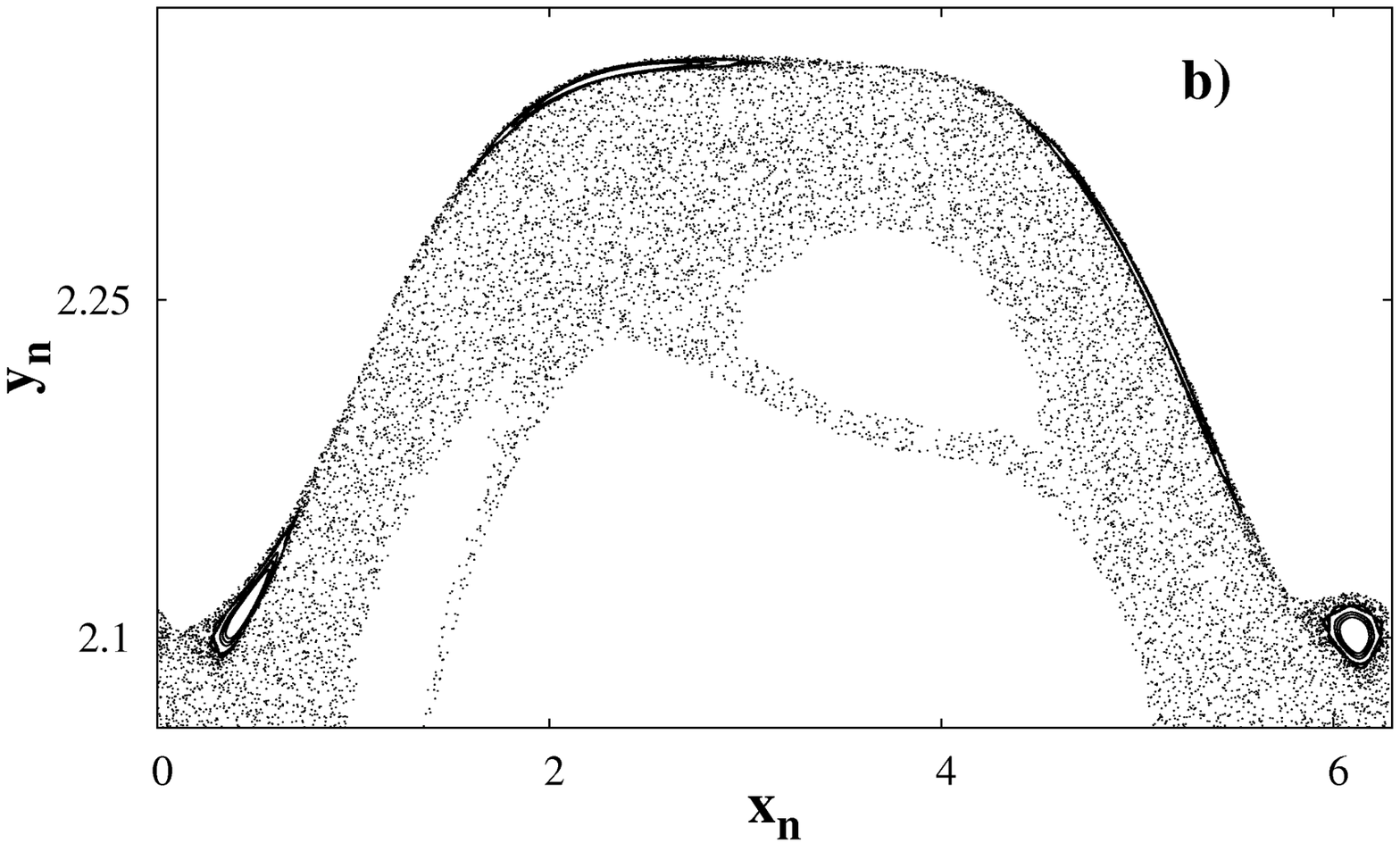}}
\caption{(a) Two $C_\mathrm{WB}^{4:1}$ western ballistic period-4 UPOs
at $\varepsilon=0.0785$ with initial
positions of 4 unstable periodic trajectories belonging to each of them.
The $e$ and $h$ orbits were born from the elliptic and hyperbolic
points of the northern ballistic resonance $4:1$,
respectively.
(b) The Poincar{\'e} section at $\varepsilon=0.005$ with
the northern ballistic resonance $4:1$ consisting of 4 ballistic islands
along the northern border of the separatrix layer.}
\label{WB}
\end{figure}

The shortest ones among all the period-4 UPOs are western ballistic UPOs.
The particles belonging to the $C_\mathrm{WB}^{4:1}$
class move in a periodic way
to the west along such an orbit in the northern separatrix
layer which appears between the northern (C) and (P) regions in
Fig.~\ref{phasespace} as a result of the perturbation.
With the help of the RM in Fig.~\ref{Rmap}a, we located two
$C_\mathrm{WB}^{4:1}$ orbits with initial positions of 4 periodic trajectories
belonging to each of them. In order to track out the origin
of the $e$ and $h$ orbits, shown in Fig.~\ref{WB}a
at $\varepsilon=0.0785$, we decrease the value of the perturbation
amplitude $\varepsilon$, compute the corresponding period-4 RMs,
locate the $C_\mathrm{WB}^{4:1}$ orbits and measure their length.

\begin{figure}[!tpb]
\centerline{\includegraphics[width=0.49\textwidth]{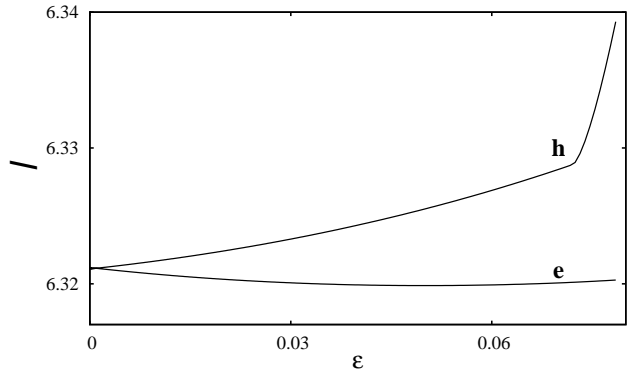}}
\caption{Dependence of the length $l$ of the $e$ and $h$
$C_\mathrm{WB}^{4:1}$ western ballistic UPOs on the perturbation amplitude
$\varepsilon$.}
\label{WBL}
\end{figure}

The result may be resumed as follows. The western ballistic
resonance $4:1$ with one elliptic and one hyperbolic points appears
under a perturbation with a very small value of $\varepsilon$.
On the Poincar{\'e} section in Fig.~\ref{WB}b at $\varepsilon=0.005$,
it is manifested as 4 ballistic islands along the northern border
of the separatrix layer. Two of the islands are so thin that they
are hardly visible in the figure.
With increasing $\varepsilon$, the size of the resonance
decreases and at the critical value $\varepsilon\approx 0.016$
it vanishes (see Fig.~\ref{Rmap}b where there are no signs of that
resonance at $\varepsilon=0.0785$).
The orbit, associated with the elliptic fixed point of the western resonance $4:1$,
loses its stability and bifurcates into the $C_\mathrm{WB}^{4:1}$ UPO
of period-4 which we denote by the symbol $e$.
Its length practically does not change with increasing $\varepsilon$
(see Fig.~\ref{WBL}).
The length $l$ of the $h$ orbit, associated with the hyperbolic point of
the that resonance, changes dramatically at the point of
bifurcation $\varepsilon\approx 0.0715$ (see Fig.~\ref{WBL})
increasing fastly after this point because of appearing a meander and a loop
on the $h$ orbit nearby the SO.

\subsection{$C_\mathrm{EB}^{4:1}$ class: eastern ballistic UPOs
associated with the $4:1$ eastern ballistic resonance}

\begin{figure}[!tpb]
\centerline{\includegraphics[width=0.49\textwidth]{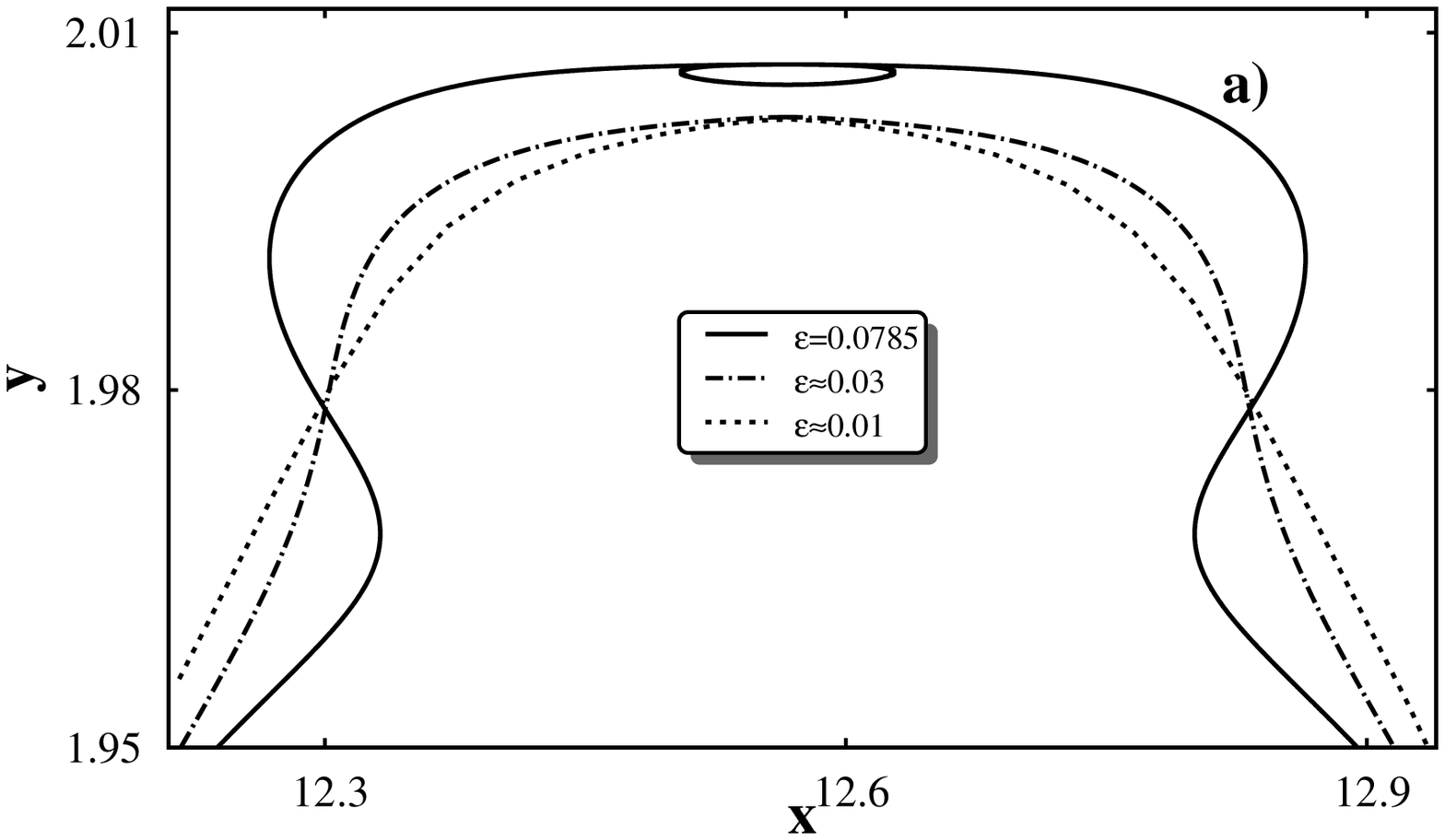}
\includegraphics[width=0.49\textwidth]{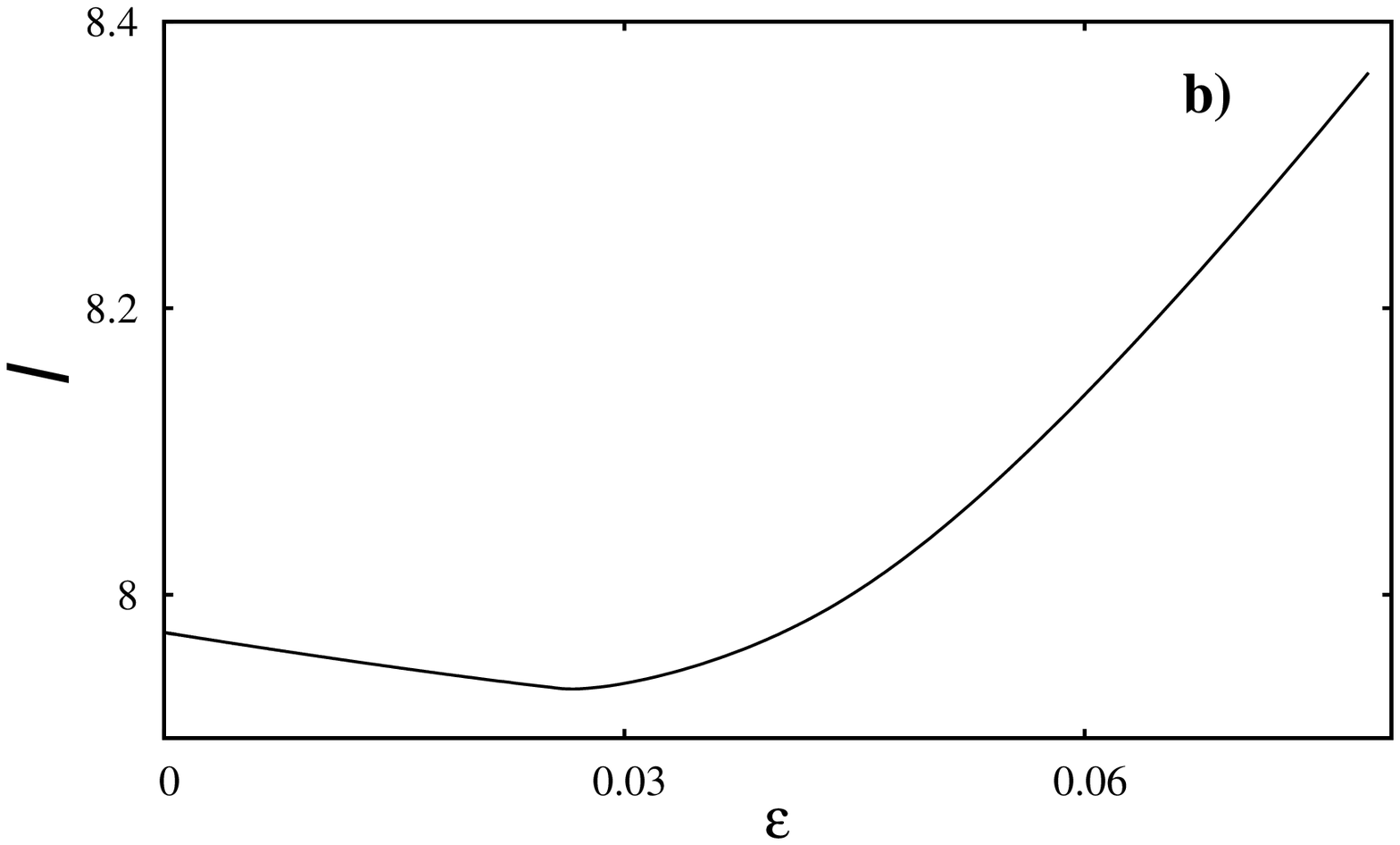}}
\caption{(a) Metamorphoses of the $h$ $C_\mathrm{EB}^{4:1}$ eastern ballistic
UPO as the perturbation amplitude $\varepsilon$ changes:
solid line, $\varepsilon=0.0785$; dashed line, $\varepsilon=0.03$;
and dotted line, $\varepsilon=0.01$. (b) Dependence of the
length $l$ of the $h$ $C_\mathrm{EB}^{4:1}$ orbit on the perturbation
amplitude $\varepsilon$.}
\label{EB}
\end{figure}

The eastern ballistic UPOs of period-4 lie in the southern separatrix
layer between the (J) and (C) regions (see Fig.~\ref{phasespace})
to the north from the jet. Particles move along such
orbits to the east in a periodic way. The origin and
bifurcations of the $C_\mathrm{EB}^{4:1}$ orbits are similar to the
western ballistic ones. They appear from elliptic and
hyperbolic points of the eastern ballistic resonance $4:1$
as $\varepsilon$ increases. The elliptic orbit of this
resonance loses it stability and bifurcates into the
UPO of the $e$-type. The hyperbolic orbit changes its topology,
transforming from the bell-like curve at $\varepsilon=0.01$ to the curve with
a meander and a loop nearby the SO at $\varepsilon=0.0785$
(see Fig.~\ref{EB}a). The length $l$ of the $h$ orbit
in Fig.~\ref{EB}b decreases with increasing $\varepsilon$
up to the bifurcation point $\varepsilon\approx 0.0275$,
after which it increases fastly due to the
comlexification of the orbit's form
(see Fig.~\ref{EB}a).

\subsection{$C_\mathrm R^{4:1}$ class: orbits associated with the $4:1$
rotational resonance}

\begin{figure}[!tpb]
\centerline{\includegraphics[width=0.49\textwidth]{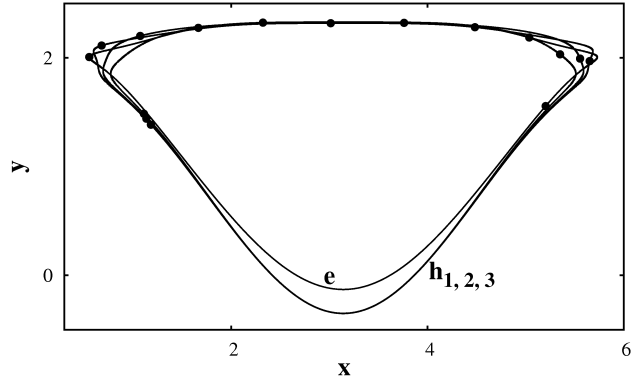}}
\caption{Four $4:1$ rotational UPOs of the $C_\mathrm R^{4:1}$ class
with initial positions of 4 unstable periodic trajectories belonging
to each of them. The $e$ and $h_{1, 2, 3}$ orbits were born from
the elliptic and hyperbolic points of the $4:1$
rotational resonance, respectively ($\varepsilon=0.0785$).}
\label{R4}
\end{figure}

\begin{figure}[!tpb]
\centerline{\includegraphics[width=0.49\textwidth]{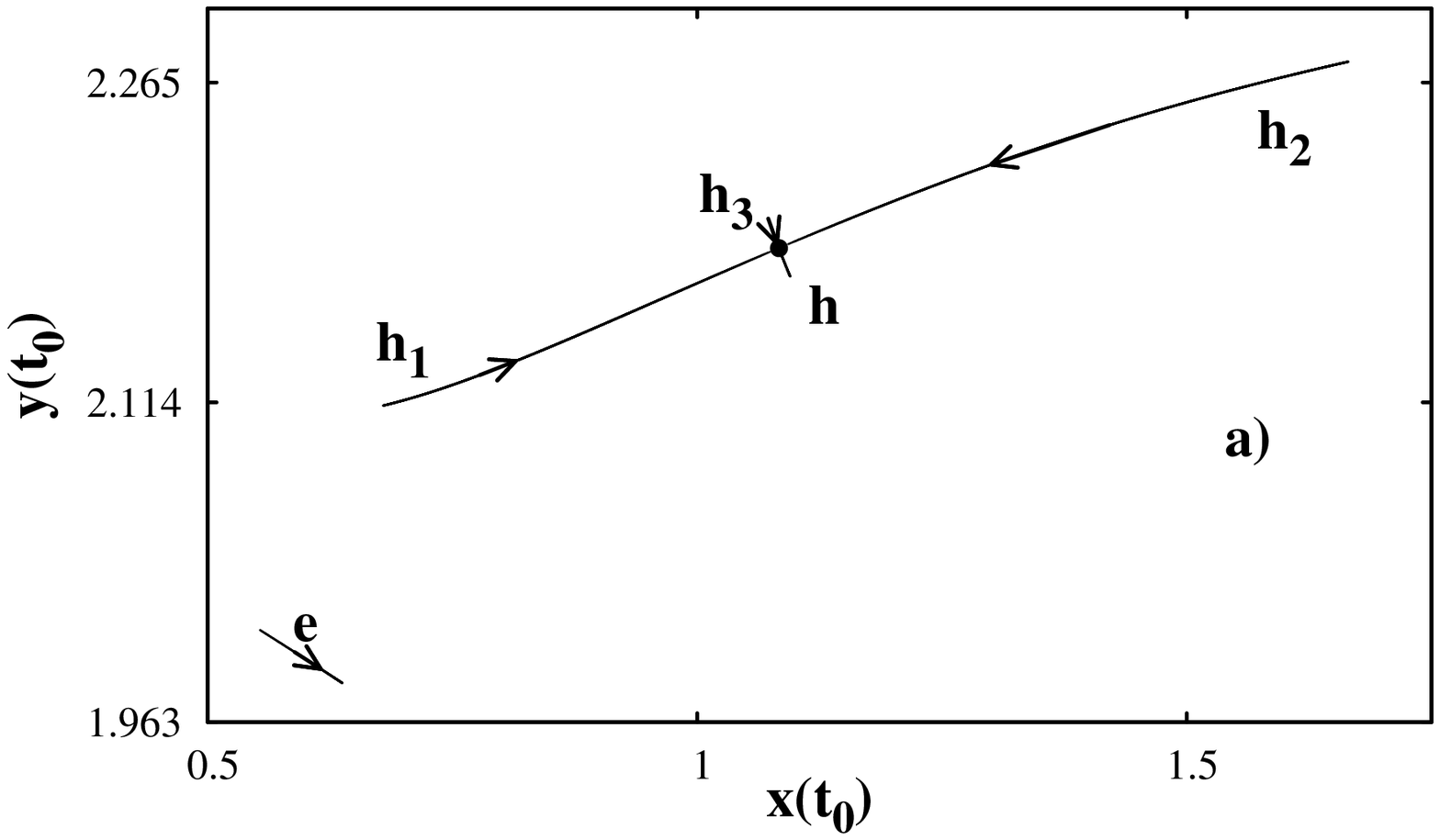}
\includegraphics[width=0.49\textwidth]{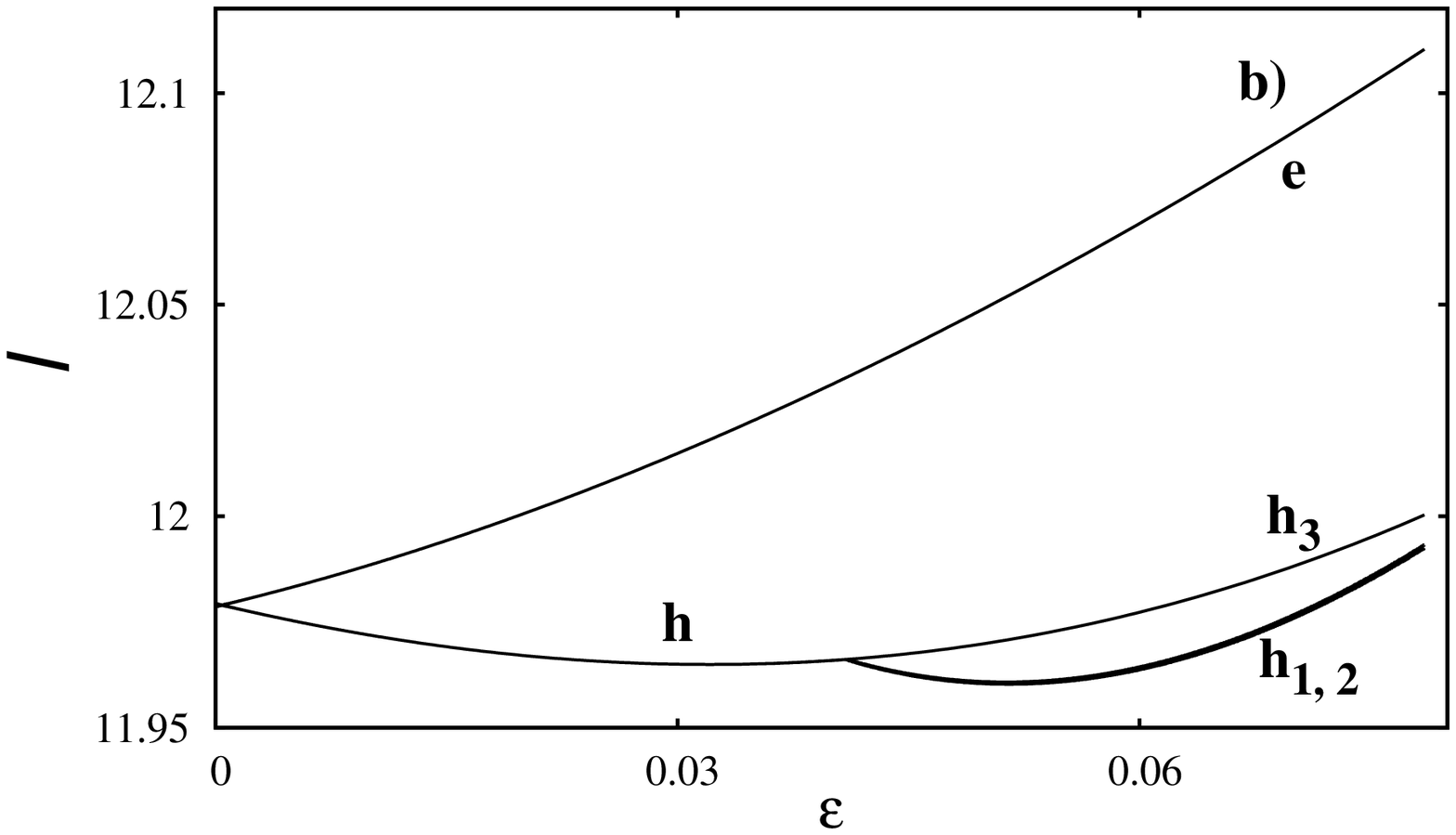}}
\caption{(a) Movement of initial positions $(x(t_0), y(t_0))$ of the
rotational $4:1$ UPOs with decreasing the perturbation amplitude
is shown by arrows. The black circle is a pitchfork bifurcation point
at $\varepsilon\approx 0.040945$ where the hyperbolic point of the
resonance $h$ bifurcates into three different orbits
$h_1, h_2$ and $h_3$. (b) Bifurcation diagram $l(\varepsilon)$.}
\label{R4B}
\end{figure}

Four different orbits in the $C_\mathrm R^{4:1}$ class, located with the help
of the period-4 RM, are shown in Fig.~\ref{R4} at
$\varepsilon=0.0785$ along with initial positions of 4 unstable periodic
trajectories on each of them.
The corresponding fluid particles rotate in the same frame along
closed curves. The genesis of the $C_\mathrm R^{4:1}$ orbits is the following.
A $4:1$ rotational resonance appears under a small perturbation
(\ref{perturb}). As the amplitude $\varepsilon$ increases, its elliptic point
loses stability and bifurcates at $\varepsilon\approx 0.003$ into the UPO
which we denote by the symbol $e$.
The hyperbolic orbit of the $4:1$ resonance $h$ bifurcates
at $\varepsilon\approx 0.040945$ into two hyperbolic orbits, $h_1$, $h_2$
and an elliptic orbit $h_3$ in the center of the rotational $4:1$ resonance.
It is a pitchfork bifurcation. In Fig.~\ref{R4B}a
we show by arrows a movement of initial positions of the
$C_\mathrm R^{4:1}$ orbits as $\varepsilon$ decreases.
The pitchfork bifurcation point is shown as a black circle.
As $\varepsilon$ increases further, the elliptic orbit $h_3$ loses
its stability and becomes a hyperbolic $C_\mathrm R^{4:1}$ orbit.

\subsection{$C_\mathrm R^{2:1}$ class: orbits associated with the $2:1$
rotational resonance}

\begin{figure}[!tpb]
\centerline{\includegraphics[width=0.49\textwidth]{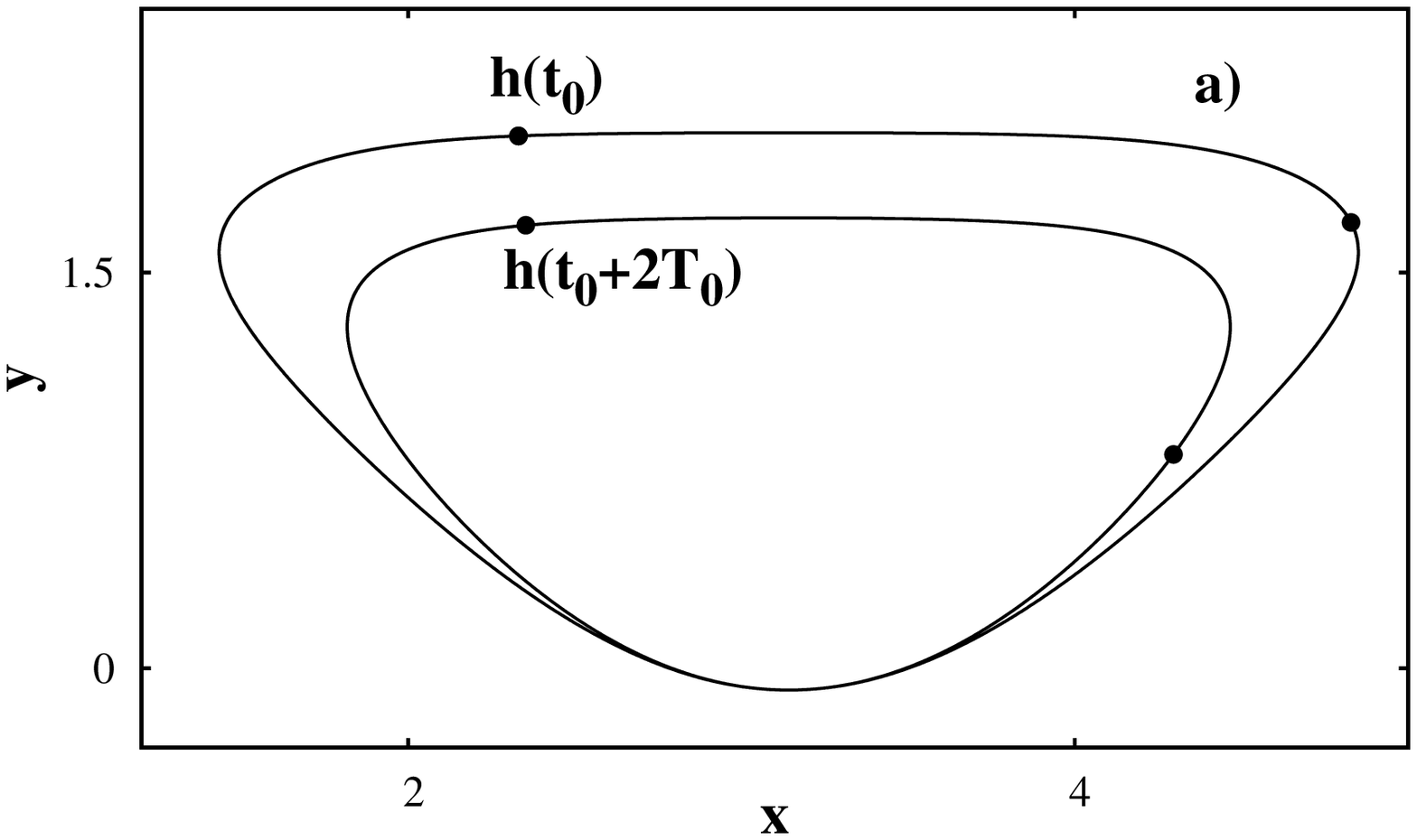}
\includegraphics[width=0.49\textwidth]{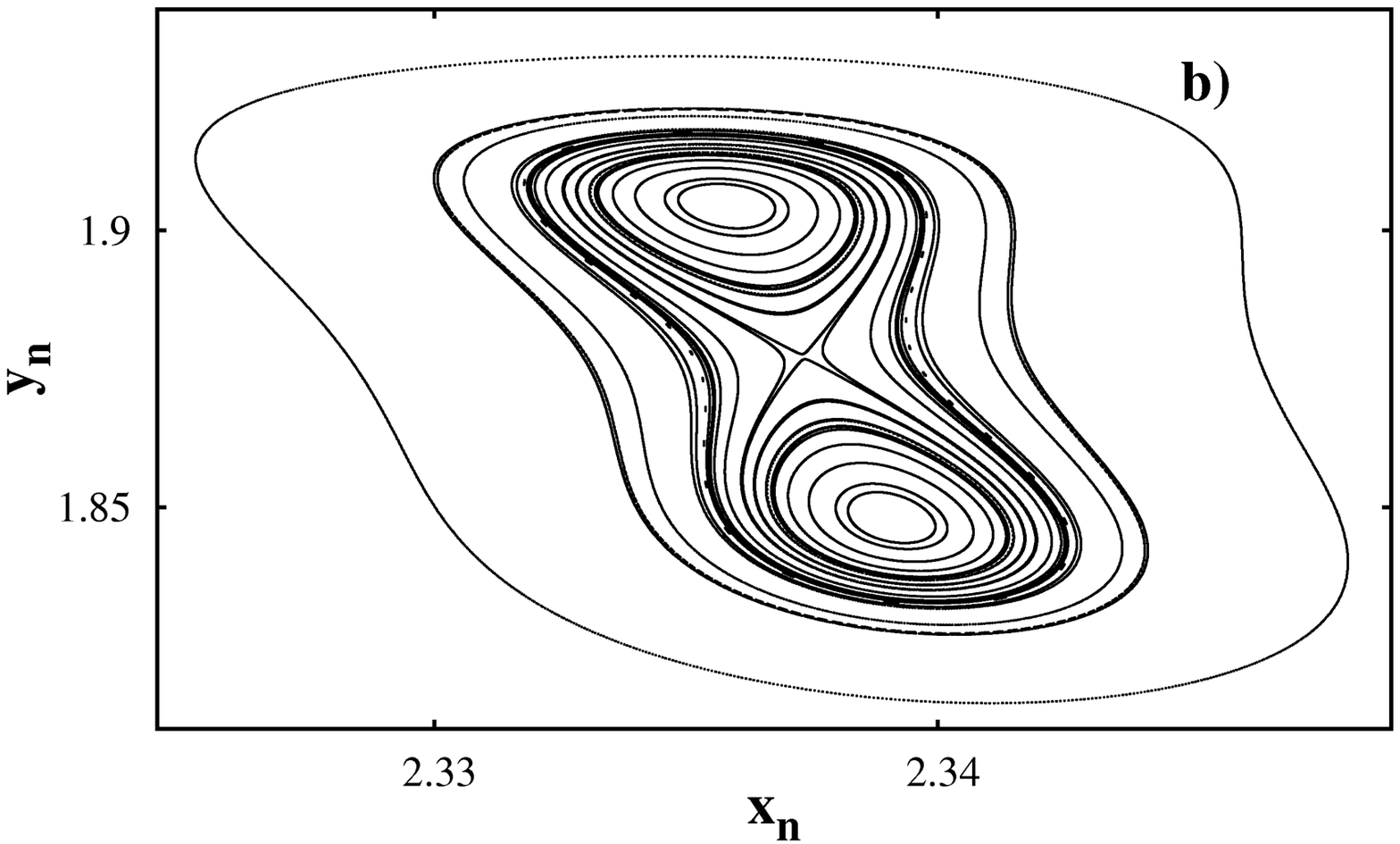}}
\caption{(a) $2:1$ rotational UPOs with initial positions of 4 unstable
periodic trajectories ($\varepsilon=0.0785$).
(b) Poincar{\'e} section of the $2:1$ rotational resonance
at $\varepsilon=0.0668$ (just after the period-2 bifurcation).}
\label{R2}
\end{figure}

\begin{figure}[!tpb]
\centerline{\includegraphics[width=0.49\textwidth]{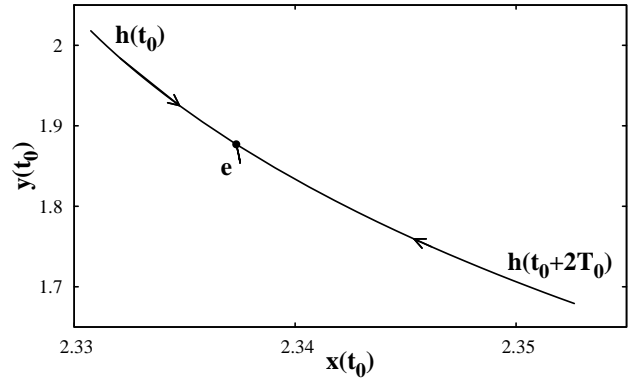}}
\caption{Emergency of the doubly-branched UPO, shown in Fig.~\ref{R2}a,
from the elliptic orbit of the $2:1$ rotational resonance at the bifurcation
point (black circle). Movement of fixed initial points
on the upper $h(t_0)$ and lower $h(t_0+2T_0)$ branches
to the bifurcation point with decreasing $\varepsilon$ is shown.}
\label{R2B}
\end{figure}

The class $C_R^{2:1}$ consists of two rotational UPOs, associated
with the $2:1$ rotational resonance, with elliptic and hyperbolic orbits
and two periodic trajectories on each of them. The resonance
appears under a small perturbation. At $\varepsilon\approx 0.0665$,
the elliptic orbit of this resonance undergoes a period-2 bifurcation
into a period-4 elliptic orbit
with 4 trajectories and a period-2 hyperbolic orbit with 2 trajectories.
The latter is shown in Fig.~\ref{R2}a at $\varepsilon=0.0785$.
The Poincar{\'e} section of the $2:1$ rotational resonance at
$\varepsilon=0.0668$ (just after the bifurcation) is shown in Fig.~\ref{R2}b.
By further increasing $\varepsilon$,
we see that the resonance $2:1$ vanishes, and the period-4 elliptic
orbit loses its stability and bifurcates into a period-4 UPO.
To clarify this bifurcation we fix a point at the moment
$t_0$ on the upper branch of the orbit in Fig.~\ref{R2}a
and another point on the lower branch at $t=t_0+2T_0$ and scan their
positions when decreasing the perturbation amplitude $\varepsilon$.
Figure~\ref{R2B} demonstrates clearly that the points
move to the position of the elliptic orbit of the $2:1$ resonance
and merge with it at the bifurcation point.

\subsection{$C_\mathrm {RB}^{4:1}$ class: rotation-ballistic UPOs
associated with the ro\-ta\-tional-ballistic resonance}

\begin{figure}[!tpb]
\centerline{\includegraphics[width=0.49\textwidth]{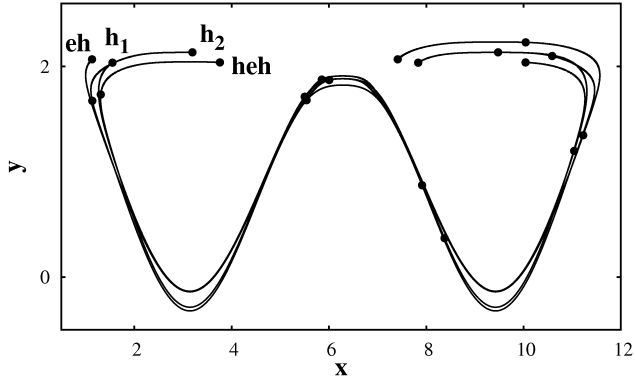}}
\caption{Four rotational-ballistic period-4 UPOs with
their initial positions in the phase space $eh$, $h_1$, $h_2$ and $heh$.
The other points are the initial positions of the corresponding
trajectories.}
\label{RB}
\end{figure}

The class $C_\mathrm{RB}^{4:1}$ consists of four period-4 UPOs shown in Fig.~\ref{RB}a.
We call those orbits as rotational-ballistic (RB) ones because the
corresponding
particles begin to move to the west in one frame, then turn to the east and
travel in the southern separatrix layer to the next frame, fulfill one
turnover in this frame and repeat their motion to the east. The
genesis of the $C_\mathrm{RB}^{4:1}$ orbits differs from the genesis of the
other classes of period-4 UPOs. Each of the resonances,
associated with $C_\mathrm{RW}^{4:1}$, $C_\mathrm{RE}^{4:1}$,
$C_\mathrm R^{4:1}$ and $C_\mathrm R^{2:1}$,
appears under an infinitely small perturbation amplitude $\varepsilon$.
Rotation-ballistic resonances of period-m and corresponding orbits cannot
in principle appear in the flow below some critical value of $\varepsilon$
(which depends on $m$)
because the width of the stochastic layers (which increases with
increasing $\varepsilon$) should be large enough in order that
particles would have enough time to travel the corresponding
distance to the east and west.

\begin{figure}[!tpb]
\centerline{\includegraphics[width=0.49\textwidth]{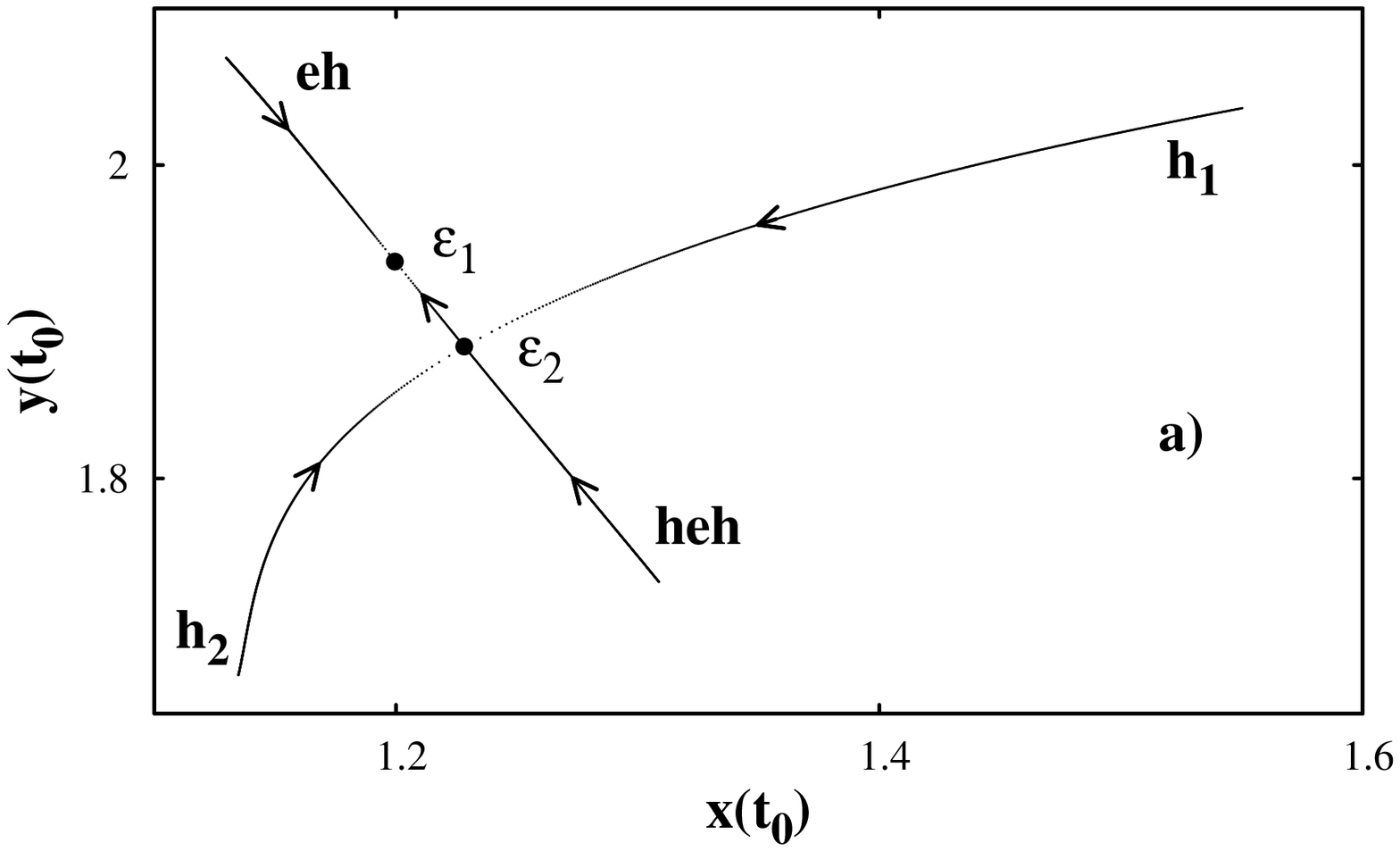}
\includegraphics[width=0.49\textwidth]{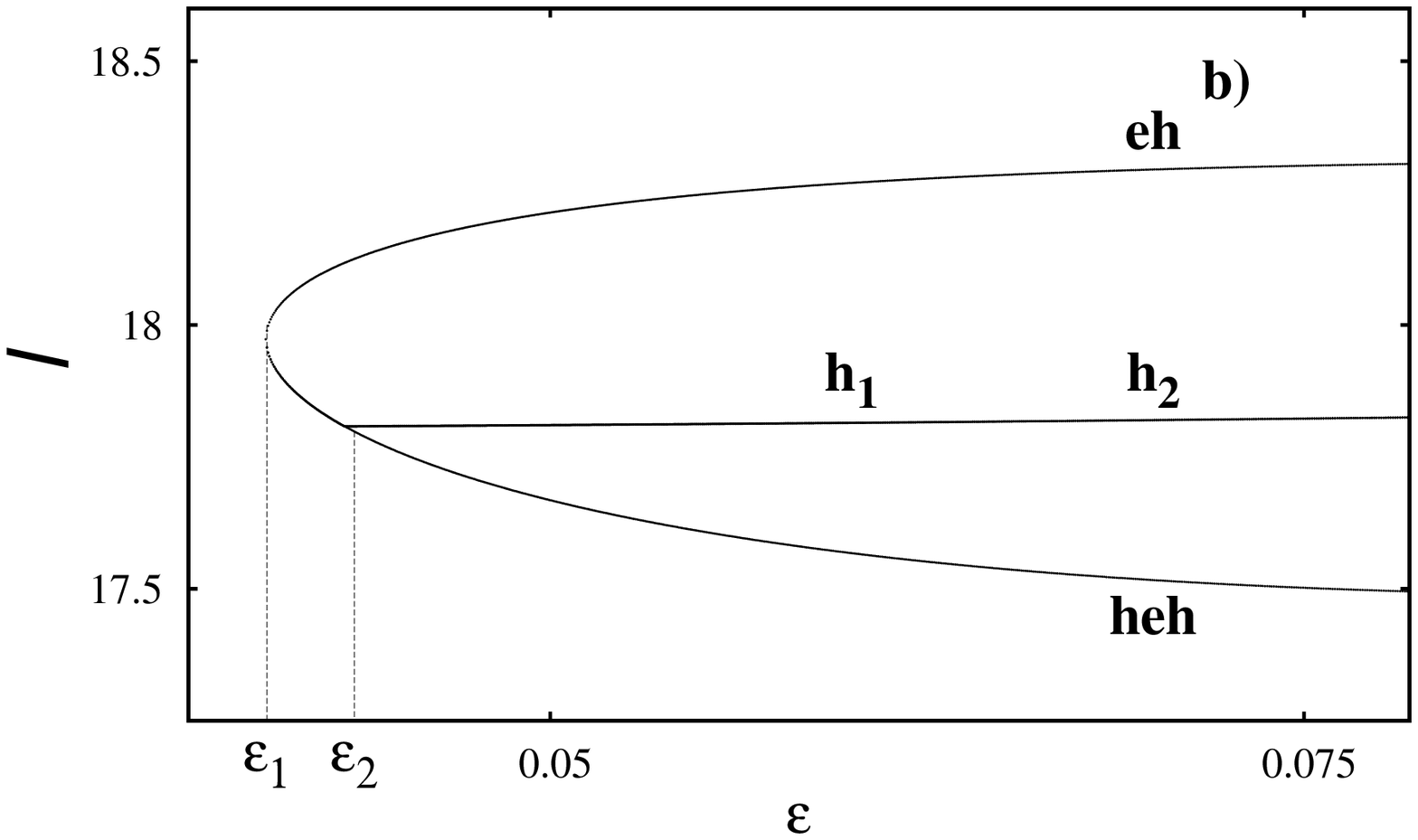}}
\caption{(a) Arrows show movement of the initial positions of the
$C_\mathrm{RB}^{4:1}$ orbits with decreasing the perturbation amplitude
$\varepsilon$.
At the point $\varepsilon=\varepsilon_1$, there appears a RB $4:1$
resonance with the elliptic orbit $eh$ which loses its stability with
increasing $\varepsilon$. The hyperbolic orbit of the RB resonance
bifurcates at $\varepsilon=\varepsilon_2>\varepsilon_1$ into two
hyperbolic period-4 UPOs $h_1$ and $h_2$ and the elliptic orbit
$heh$ which loses its stability with further increasing $\varepsilon$.
(b) Bifurcation diagram $l(\varepsilon)$.}
\label{RBB}
\end{figure}

\begin{figure}[!tpb]
\centerline{\includegraphics[width=0.49\textwidth]{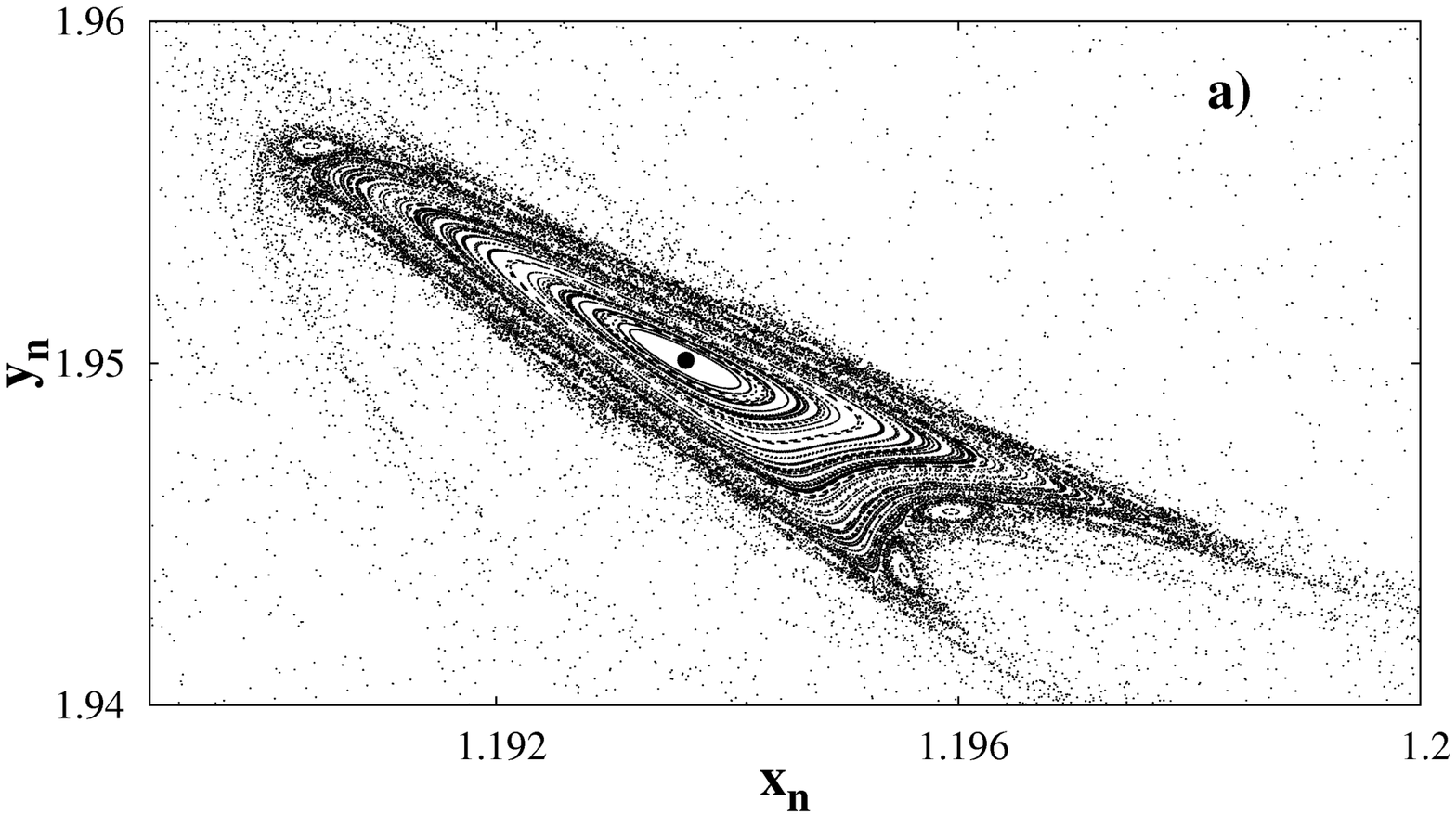}
\includegraphics[width=0.49\textwidth]{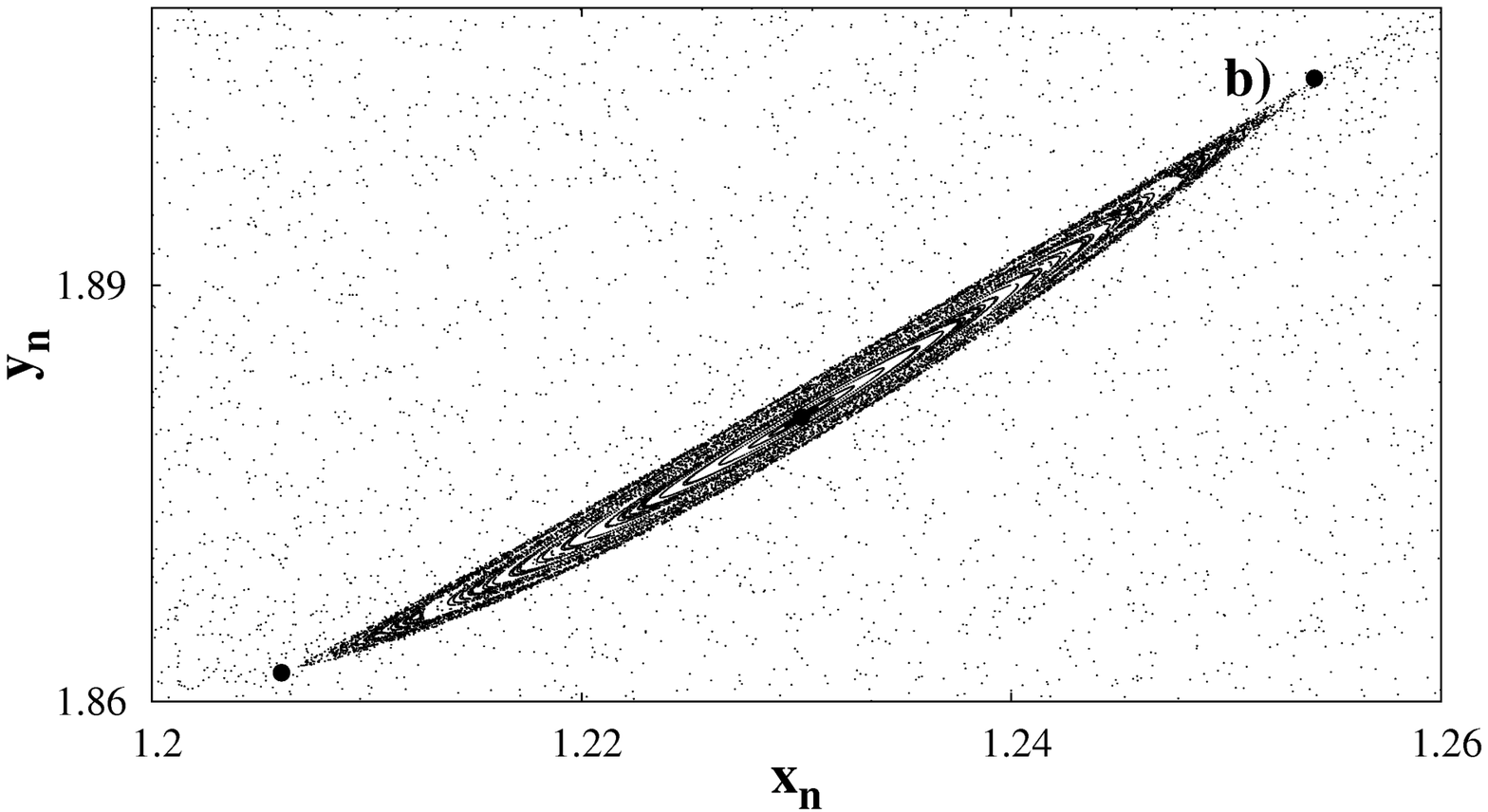}}
\caption{(a) One of the stability islands of the RB $4:1$ resonance
appearing after the first bifurcation at $\varepsilon=0.04072\gtrsim \varepsilon_1$ with the elliptic
point $eh$ in its center. (b) One of the stability islands of the
RB $4:1$ resonance appearing after a pitchfork bifurcation at
$\varepsilon=0.0435 \gtrsim \varepsilon_2$
with the elliptic point $heh$ in its center.}
\label{ab}
\end{figure}

The genesis and evolution of the $C_\mathrm {RB}^{4:1}$ UPOs are shown
schematically on the bifurcation diagrams in Fig.~\ref{RBB}.
In Fig.~\ref{RBB}a we demonstrate movement in the phase space of
initial positions of the $C_\mathrm {RB}^{4:1}$ orbits with
decreasing the perturbation amplitude $\varepsilon$.
The dependence of the lengths of those orbits on $\varepsilon$
is shown in Fig.~\ref{RBB}b.
The rotational-ballistic $4:1$ resonance appears at a critical value of the
perturbation amplitude $\varepsilon=\varepsilon_1\approx 0.040715$
and manifests itself as 4 small islands of stability on the corresponding
Poincar{\'e} section. One of these islands is shown in Fig~\ref{ab}a.
If $\varepsilon$ increases further, the elliptic orbit $eh$
in the center of the RB $4:1$ resonance loses its stability (it was
checked by computing an accumulated error) and becomes a hyperbolic
RB period-4 orbit. The hyperbolic orbit $h$ of the RB resonance
at $\varepsilon=\varepsilon_2\approx 0.0433$ undergoes a pitchfork
bifurcation into two hyperbolic RB period-4 orbits $h_1$ and $h_2$
and the elliptic orbit $heh$ in the center of the new period-4 RB
resonance one of whose stability islands is shown in Fig.~\ref{ab}b.
Under further increasing $\varepsilon$, the elliptic orbit $heh$
loses its stability and becomes a hyperbolic RB period-4 orbit.

\section{Conclusion}
We have developed a method to track unstable periodic orbits
based on computing and analyzing local minima of the distance function
$d(x(t_0),y(t_0))$. The method is not restricted to our simple Hamiltonian
system with a periodic perturbation and is applicable to a variety of chaotic
systems. The results, obtained with our specific model of a meandering jet,
can be resumed as follows.

We detected and located the UPOs of periods 1 and 4. Varying the
perturbation amplitude, we have found bifurcations of the period-4 orbits,
the UPOs of the lowest period with non-trivial origin.
Those orbits have been grouped into five classes by their origin and
bifurcations.
$C_\mathrm{WB}^{4:1}$ ($C_\mathrm{EB}^{4:1}$) class consists of the western (eastern)
ballistic UPOs associated with the $4:1$ western (eastern) ballistic resonance.
$C_\mathrm{R}^{4:1}$ ($C_\mathrm{R}^{2:1}$) class consists of the rotational UPOs associated
with the $4:1$ ($2:1$) rotational resonance. $C_\mathrm{RB}^{4:1}$ class
consists of specific rotational-ballistic UPOs associated with $4:1$
rotational-ballistic resonance. We would like to stress the following.
Rotational and ballistic resonant islands are well known objects
in the phase space of Hamiltonian systems \cite{Z05}. As far as
we know, rotation-ballistic resonant islands of stability have
not been found before. They are expected to appear as well in other jet-flow
models (kinematic and dynamic ones) under specific values of control parameters,
and it would be interesting to search for them in the models introduced
in Refs.~\cite{S92,DW96,PBUZ06,SamWig2006,UBP07, WK89, SMS89, P91, Yang96, DM93, SHS93, NS97, CLVZ99, YPJ02}.
It is expected that boundaries of the rotational-ballistic islands are
specific dynamical traps that should affect transport and statistical
properties of passive particles and may be treated by the methods developed in
\cite{UBP07}.

In addition, by linearizing the
advection equations, we have studied the properties of the period-1 saddle
orbits playing  a crucial role in chaotic transport and mixing of passive
particles. The UPOs with larger values of the period can be
classified into three big groups: ballistic, rotational and
rotation-ballistic ones. The origin and bifurcations of the orbits
in each class can be studied in a similar way, but it may require larger
computational efforts.

The results obtained do not depend critically on the exact form of the
streamfunction and chosen values of the control parameters. The UPOs in
other kinematic and dynamical models of geophysical jets, known
in the literature
\cite{WK89, SMS89, P91, Yang96, DM93, SHS93, NS97, CLVZ99, YPJ02},
can be detected, located and classified in a similar way.
The UPOs form the skeleton for chaotic advective mixing
and transport in fluid flows, and the knowledge of them (at least,
lower-period ones) allows us to analyze complex, albeit rather
regular, stretching and folding structures in the flows.

\section*{Acknowledgments}
The work was supported by the Russian Foundation
for Basic Research (Grant  no.~06-05-96032), by the Program
``Mathematical Methods in  Nonlinear Dynamics'' of the Russian
Academy of Sciences, and by the Program for Basic Research of
the Far Eastern Division of the Russian Academy of
Sciences.

\end{document}